# Catalan Solids Derived From 3D-Root Systems and Quaternions

Mehmet Koca[1] and Nazife Ozdes Koca[2]

Department of Physics, College of Science, Sultan Qaboos University

P.O. Box 36, Al-Khoud, 123 Muscat, Sultanate of Oman

Ramazan Koç[3]

Department of Physics, Gaziantep University, 27310, Gaziantep, Turkey

**Abstract**

Catalan Solids are the duals of the Archimedean solids, vertices of which can be obtained from the Coxeter-Dynkin diagrams $A_3$, $B_3$ and $H_3$ whose simple roots can be represented by quaternions. The respective Weyl groups $W(A_3)$, $W(B_3)$ and $W(H_3)$ acting on the highest weights generate the orbits corresponding to the solids possessing these symmetries. Vertices of the Platonic and Archimedean solids result as the orbits derived from fundamental weights. The Platonic solids are dual to each others however duals of the Archimedean solids are the Catalan solids whose vertices can be written as the union of the orbits, up to some scale factors, obtained by applying the above Weyl groups on the fundamental highest weights $(100)$, $(010)$, $(001)$ for each diagram. The faces are represented by the orbits derived from the weights $(010)$, $(110)$, $(101)$, $(011)$ and $(111)$ which correspond to the vertices of the Archimedean solids. Representations of the Weyl groups $W(A_3)$, $W(B_3)$ and $W(H_3)$ by the quaternions simplify the calculations with no reference to the computer calculations.

[1] electronic-mail: kocam@squ.edu.om

[2] electronic-mail: nazife@squ.edu.om

[3] electronic-mail: koc@gantep.edu.tr

## 1. Introduction

Discovery of the Platonic solids; tetrahedron, cube, octahedron, icosahedron and dodecahedron dates back to the people of Scotland lived 1000 years earlier than the ancient Greeks and the models curved on the stones are now kept in the Ashmolean Museum at Oxford [1].

It took nearly another century after Plato's association of tetrahedron with fire, cube with earth, air with octahedron, and water with icosahedron that Archimedes discovered the semi-regular convex solids. However, several centuries passed to rediscover them by the renaissance mathematicians. Finally Kepler completed the work in 1620 by introducing prisms and anti-prisms as well as four regular non-convex polyhedra, now known as the Kepler-Poinsot polyhedra. Construction of the dual solids of the Archimedean solids was completed in 1865 by Catalan [2] nearly two centuries after Kepler. Extension of the platonic solids to 4D dimension has been made in 1855 by L. Schlaffli [3] and their generalizations to higher dimensions in 1900 by T.Gosset [4]. Further important contributions are made by W. A. Wythoff [5] among many others and in particular by the contemporary mathematician H.S.M. Coxeter [6] and J.H. Conway [7].

The 3D and 4D dimensional convex polytopes single out as compared to the polytopes in higher dimensions. The number of Platonic solids is five in 3D and there exist six regular polytopes in 4D on the contrary to the higher dimensional cases where we have only three platonic solids.

Let $G$ be the root system of rank n. The orbit $O(a_1 a_2 ... a_n) \equiv W(G)(a_1 a_2 ... a_n)$ defines a polytope (non-regular in general) possessing the $W(G)$ symmetry, where $\Lambda = (a_1 a_2 ... a_n)$ is the highest weight with $a_i$ (i=1,2,...,n) non-negative integers [8] and $W(G)$ is the Coxeter-Weyl group. The orbits $O(10...0) = W(A_n)(10...0)$ and $O(00...1) = W(A_n)(00...1)$ each defines an (n+1)-cell (or n-simplex). They can be transformed to each other because of the Dynkin-diagram symmetry of the root system of $A_n$. They are said to be self-dual. The orbits $O(10...0) = W(B_n)(10...0)$ and $O(00...1) = W(B_n)(00...1)$ represent super octahedron and super cube in *n*-dimensions which are dual to each other. These are the only Platonic polytopes existing in arbitrary dimensions. In 3D, in addition to the self-dual tetrahedron, octahedron and cube there are two more polyhedra, the icosahedron and dodecahedron described by the Coxeter diagram $H_3$. In 4D, in addition to the Platonic Polytopes described by the symmetries of $W(A_4)$ and $W(B_4)$, we have 120-cell and 600-cell described by the Coxeter group $W(H_4)$ and the 24-cell whose vertices described by the either orbits $O(1000) = W(F_4)(1000)$ and $O(0001) = W(F_4)(0001)$. The 24-cell is said to be self-dual because the $F_4$ diagram is invariant under the Dynkin-diagram symmetry [9, 10].

Having said these general statements for the Platonic polytopes in arbitrary dimensions, we return back to the semi-regular polyhedra in 3D dimensions. In ref. [11] we have studied, in detail, the Platonic and the Archimedean solids employing the above technique where the simple roots are described by quaternions. Consequently, the Coxeter-Weyl groups were expressed using the binary tetrahedral, binary octahedral and the binary icosahedral subgroups of quaternions. In this paper we construct the vertices of the Catalan solids, duals of the Archimedean solids, using the same technique described in the paper [11]. In section 2 we set up the general frame by summarizing the technique of ref. [11] and construct the vertices of the first Catalan solid invariant under the tetrahedral group $W(A_3) \approx T_d$. Section 3 is devoted to the discussion of the Catalan solids possessing the octahedral symmetry $W(B_3) \approx O_h$. The Catalan solids described by the icosahedral symmetry $W(H_3) \approx I_h$ will be studied in section 4. Section 5 is based on the concluding remarks regarding the use of the Catalan solids in the description of viral structures and other applications.

2. **Quaternionic descriptions of the rank-3 Coxeter-Dynkin Diagrams**

Let $q = q_0 + q_i e_i$, $(i = 1,2,3)$ be a real unit quaternion with its conjugate defined by $\bar{q} = q_0 - q_i e_i$ and the norm $q\bar{q} = \bar{q}q = 1$. Here the quaternionic imaginary units satisfy the relations

$$e_i e_j = -\delta_{ij} + \varepsilon_{ijk} e_k, \ (i, j, k = 1,2,3) \tag{1}$$

where $\delta_{ij}$ and $\varepsilon_{ijk}$ are the Kronecker and Levi-Civita symbols and summation over the repeated indices is implicit. With the definition of the scalar product

$$(p, q) = \frac{1}{2}(\bar{p}q + \bar{q}p) = \frac{1}{2}(p\bar{q} + q\bar{p}), \tag{2}$$

quaternions generate the four-dimensional Euclidean space. The group of quaternions is isomorphic to $SU(2)$ which is the double cover of the proper rotation group $SO(3)$. Its finite subgroups are the cyclic groups of order $n$, the dicyclic groups of order $2n$, binary tetrahedral group $T$ of order 24, the binary octahedral group $O$ of order 48 and finally the binary icosahedral group $I$ of order 120 [12]. To set the scene let us define the 4D orthogonal transformations in terms of quaternions. Let $p$ and $q$ be two arbitrary unit quaternions and $r$ represent any quaternion. Then the $O(4)$ transformations can be defined by

$$[p, q] : r \to r' = prq, \ [p, q]^* : r \to r'' = p\bar{r}q. \tag{3}$$

When the unit quaternions $p$ and $q$ take values from the binary icosahedral group $I$, the set of elements

$$W(H_4) = \{[p, q] \oplus [p, q]^*\} \tag{4}$$

represents the Coxeter group of order 14,400, the symmetry of the 120-cell and the 600-cell [13]. When $q = \bar{p}$ is substituted in (4) we obtain the Coxeter group $W(H_3)$ representing the icosahedral symmetry $I_h$ of order 120,

$$W(H_3) = \{[p, \bar{p}] \oplus [p, \bar{p}]^*\} \approx A_5 \times C_2, p \in I \tag{5}$$

where even permutations of five letters $A_5 = \{[p, \bar{p}], p \in I\}$ is the proper icosahedral group without inversion and the generator $[1,1]^*$ representing the inversion in three dimensions generates the cyclic group $C_2$.

We can represent the elements of the binary octahedral group $O$ as follows

$$O = T \oplus T'. \tag{6}$$

Here $T$ represents the binary tetrahedral group given by

$$T = \{ \pm 1, \pm e_1, \pm e_2, \pm e_3, \frac{1}{2}(\pm 1 \pm e_1 \pm e_2 \pm e_3)\} \qquad (7)$$

which is also representing the quaternionic vertices of the polytope 24-cell. The dual polytope is represented by the quaternions

$$T' = \{\tfrac{1}{\sqrt{2}}(\pm 1 \pm e_1), \tfrac{1}{\sqrt{2}}(\pm e_2 \pm e_3), \tfrac{1}{\sqrt{2}}(\pm 1 \pm e_2), \tfrac{1}{\sqrt{2}}(\pm e_3 \pm e_1), \tfrac{1}{\sqrt{2}}(\pm 1 \pm e_3), \tfrac{1}{\sqrt{2}}(\pm e_1 \pm e_2)\}. \qquad (8)$$

Let $p, q \in O$ be arbitrary elements of the binary octahedral group then the set of elements

$$Aut(F_4) \approx W(F_4) : C_2 = \{[p,q] \oplus [p,q]^*\} \qquad (9)$$

is the extension of the Coxeter-Weyl group $W(F_4)$ by the diagram symmetry [9]. The Weyl group $W(F_4)$ is the symmetry group of the 24-cell, the vertices of which can be represented either by the set $T$ or $T'$. The group $C_2$ in $Aut(F_4)$ transforms $T$ and $T'$ to each other where the group $C_2$ represents the Dynkin diagram symmetry of the $F_4$ diagram. If $p, q \in T$ then the group $W(D_4) : C_3 = \{[p,q] \oplus [p,q]^*\}$ of order 576 is a maximal subgroup in both groups $W(F_4)$ and $W(H_4)$ and represent the symmetry group of the *snub 24-cell* [14]. The group $W(F_4)$ has several subgroups acting in 3D space. The Weyl group $W(A_3) \approx S_4$ can be represented by the set of elements

$$W(A_3) = \{[p, \bar{p}] \oplus [t, \bar{t}]^*\}, \; p \in T, t \in T'. \qquad (10)$$

It is a subgroup of both groups $W(H_4)$ and $W(F_4)$. Another subgroup of the group $W(F_4)$ is the octahedral group $O_h \approx W(B_3)$ described by the set of elements

$$W(B_3) = \{[p, \bar{p}] \oplus [t, \bar{t}] \oplus [p, \bar{p}]^* \oplus [t, \bar{t}]^*\}. \qquad (11)$$

In ref. [11] we have shown that the orbit $O(a_1 a_2 a_3) = W(A_3)((a_1 a_2 a_3))$ can be written as

$$\begin{aligned}
&\pm \alpha e_1 \pm \beta e_2 \pm \gamma e_3; \pm \beta e_1 \pm \gamma e_2 \pm \alpha e_3; \pm \gamma e_1 \pm \alpha e_2 \pm \beta e_3; \\
&\pm \alpha e_1 \pm \gamma e_2 \pm \beta e_3; \pm \gamma e_1 \pm \beta e_2 \pm \alpha e_3; \pm \beta e_1 \pm \alpha e_2 \pm \gamma e_3
\end{aligned} \qquad (12)$$

(even number of (-) sign)

where $\alpha = \frac{1}{2}(a_1 - a_3), \beta = \frac{1}{2}(a_1 + a_3), \gamma = \frac{1}{2}(a_1 + 2a_2 + a_3)$.

The orbits $O(100)$ and $O(001)$ each represents a tetrahedron. The orbits $O(010)$ and $O(111)$ respectively represent octahedron and truncated octahedron which possess a larger octahedral symmetry $W(B_3) \approx Aut(A_3) \equiv W(A_3) : C_2$ and will be considered in the next section. Here we have only truncated tetrahedron, an Archimedean solid, having tetrahedral symmetry, which can be represented either by the orbit $O(110)$ or the orbit $O(011)$. The vertices of the orbit $O(110)$ is given by the set of quaternions

$$\tfrac{1}{2}(\pm e_1 \pm e_2 \pm 3e_3), \tfrac{1}{2}(\pm e_1 \pm 3e_2 \pm e_3), \tfrac{1}{2}(\pm 3e_1 \pm e_2 \pm e_3) \text{ (even number of } (-) \text{ sign)}. \qquad (13)$$

It consists of 12 vertices, 18 edges and 8 faces (4 equilateral triangles and 4 regular hexagons). The dual of this solid is the Catalan solid called *triakis tetrahedron* consisting of 8 vertices, 18 edges and 12 faces. Before we proceed further we note a very important property, namely, the order of the Weyl-Coxeter group divided by the size of the orbit, $|W(G)|/|O(a_1 a_2 ... a_n)|$ gives the order of a subgroup of the group $W(G)$ fixing one element of the orbit. In the above case it is a cyclic group $C_2$ of order 2. In the Catalan solids one face is represented by the vertex of the Archimedean solid orthogonal to this face. Therefore the symmetry fixing one vertex of the Archimedean solid is the symmetry of the face of the Catalan solid. Keeping this aspect of the symmetry of the face of the Catalan solid in mind we now describe the vertices and the faces of the *triakis tetrahedron*. A table of Catalan solids can be found in the reference [15].

Let us take one of the vertex of the truncated tetrahedron, say, the highest weight

$$q = \Lambda = (110) = (100) + (010) = \frac{1}{2}(e_1 + e_2 + e_3) + e_3. \tag{14}$$

The centers of the triangular faces of the truncated tetrahedron can be represented by a scaled copy of the set of quaternions of the orbit which represents the vertices of a tetrahedron

$$O(100) = \{\frac{1}{2}(e_1 + e_2 + e_3), \frac{1}{2}(e_1 - e_2 - e_3), \frac{1}{2}(-e_1 - e_2 + e_3), \frac{1}{2}(-e_1 + e_2 - e_3)\} \tag{15}$$

and the centers of the hexagonal faces are represented up to a scale factor by the set of quaternions of the orbit representing the dual tetrahedron

$$O(001) = \{\frac{1}{2}(-e_1 - e_2 - e_3), \frac{1}{2}(-e_1 + e_2 + e_3), \frac{1}{2}(+e_1 + e_2 - e_3), \frac{1}{2}(+e_1 - e_2 + e_3)\}. \tag{16}$$

It is not difficult to see that the line joining two quaternions from (16)

$$B = \frac{1}{2}(-e_1 + e_2 + e_3), C = \frac{1}{2}(e_1 - e_2 + e_3) \tag{17}$$

To each other is orthogonal to the vertex of the truncated tetrahedron given in (14).

The quaternion $A = \frac{1}{2}(e_1 + e_2 + e_3)$ from (15) and the quaternions $B$ and $C$ form an isosceles triangle however it is not orthogonal to the vertex of the truncated tetrahedron in (14). In order to obtain an isosceles triangle orthogonal to the vertex of (14) one can change the scale of the vertex at $A$ and redefine it as $A = \frac{\lambda}{2}(e_1 + e_2 + e_3)$. Now one can check that the isosceles triangle $ABC$ is orthogonal to the vertex in (14) provided $\lambda = \frac{3}{5} = 0.6$. Relative magnitude of these vertices is $|A|/|B| = |A|/|C| = 0.6$. We know that when acting by $W(A_3)$ on $A$ will generate the set $\lambda O(100)$ and similarly when acting on the $B$ or $C$ it will generate the set $O(001)$ then the vertices of the *triakis tetrahedron* lie on two orbits of $W(A_3)$. If we rescale the quaternions on these two orbits as the unit quaternions then the vertices of the *triakis tetrahedron* lie on two spheres of radii 0.6 and 1. To determine the dihedral angle between two adjacent isosceles triangular faces we rotate the system around the unit quaternion $\frac{1}{\sqrt{3}}(e_1+e_2+e_3)$ by the element $[\frac{1}{2}(1+e_1+e_2+e_3), \frac{1}{2}(1-e_1-e_2-e_3)]$ of the group $W(A_3)$ which leads to the isosceles triangle $ACD$ where $D=\frac{1}{2}(e_1+e_2-e_3)$. The vertex of the truncated tetrahedron orthogonal to the triangular face $ACD$ is the quaternion $q'=\frac{1}{2}(e_1+e_2+e_3)+e_1$. The obtuse angle between the quaternions $q$ and $q'$ is the dihedral angle $\theta = 129°31'16''$ between two adjacent faces of the *triakis tetrahedron*. Since for the vertex $A$ we have three isosceles triangular faces we have altogether $3\times 4 = 12$ faces of the *triakis tetrahedron* and its 8 vertices are in two orbits of the group $W(A_3)$. One orbit cannot be transformed to the other orbit by the group $W(A_3)$ therefore the *triakis tetrahedron* is not vertex transitive but it is face transitive because the faces are represented by the vertices of the truncated tetrahedron. It has 18 edges; 12 of which are of equal length, each joins one element from $O(100)$ to two elements of $O(001)$ and 6 edges of equal length joining elements of $O(001)$. The truncated tetrahedron and its dual *triakis tetrahedron* are depicted in Figure 1.

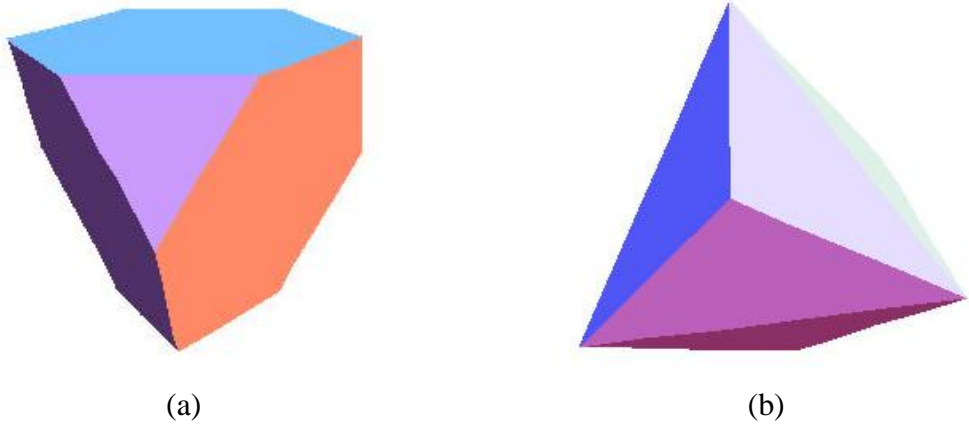

(a)          (b)

Figure 1. Truncated tetrahedron (a) and its dual *triakis tetrahedron* (b)

### 3. The Catalan solids possessing the octahedral symmetry $W(B_3)$

The Archimedean solids with the octahedral symmetry can be represented by the orbits obtained by the highest weights $(010)$ (cuboctahedron), $(110)$ (truncated octahedron), $(011)$ (truncated cube), $(101)$ (small rhombicuboctahedron) and $(111)$ (great rhombicuboctahedron) [11]. The Platonic solids, octahedron and its dual cube are represented by the orbits obtained from the highest weights $(100)$ and $(001)$ respectively. A general orbit derived from the highest weight $\Lambda = (a_1 a_2 a_3) = \alpha e_1 + \beta e_2 + \gamma e_3$ can be represented by the following 48 quaternions

$$O(a_1 a_2 a_3) = \{\pm\alpha e_1 \pm \beta e_2 \pm \gamma e_3; \pm\beta e_1 \pm \gamma e_2 \pm \alpha e_3; \pm\gamma e_1 \pm \alpha e_2 \pm \beta e_3;$$
$$\pm\alpha e_1 \pm \gamma e_2 \pm \beta e_3; \pm\gamma e_1 \pm \beta e_2 \pm \alpha e_3; \pm\beta e_1 \pm \alpha e_2 \pm \gamma e_3\}$$
(18)

Here we have $\alpha = (a_1 + a_2 + \frac{a_3}{\sqrt{2}})$, $\beta = (a_2 + \frac{a_3}{\sqrt{2}})$, $\gamma = \frac{a_3}{\sqrt{2}}$. The orbit $O(100) = \{\pm e_1, \pm e_2, \pm e_3\}$ represents the vertices of an octahedron. The orbit $O(001) = \frac{1}{\sqrt{2}}(\pm e_1 \pm e_2 \pm e_3)$ represents the vertices of the cube and the orbit $O(010) = \{(\pm e_1 \pm e_2), (\pm e_2 \pm e_3), (\pm e_3 \pm e_1)\}$ represents the vertices of a cuboctahedron. These orbits are essential for the determination of the vertices of all Catalan solids possessing the octahedral symmetry. Now we discuss the construction of the vertices of the *rhombic dodecahedron* (dual of cuboctahedron)

*3.1 Rhombic dodecahedron* (dual of cuboctahedron)

The cuboctahedron is depicted in the Figure 2(a) which shows that at one vertex two square faces and two triangular faces meet. The number of square faces is 6 and their centers are represented, up to a scale factor, by the orbit $O(100) = \{\pm e_1, \pm e_2, \pm e_3\}$ up to a scale factor. Similarly there are 8 triangular faces and their centers are represented by the orbit $O(001) = \frac{1}{\sqrt{2}}(\pm e_1 \pm e_2 \pm e_3)$ scaled by some factor. The vertex represented by the highest weight $q = (010) = e_1 + e_2$ is surrounded by two square faces whose centers, up to a scale factor, are represented by the quaternions $A = e_1$ and $C = e_2$ which belong to the orbit $O(100)$. Similarly, the centers of the triangular faces meeting at the same point can be represented, up to a scale factor, by the vertices $B = e_1 + e_2 + e_3$ and $D = e_1 + e_2 - e_3$ which belong to the orbit $O(001)$. It is clear that the lines $AC$ and $BD$ are orthogonal to the vertex $q$. However the vertex $q$ is not orthogonal to the lines $AB$ and $AD$. Keeping $A = e_1$, $C = e_2$ as they are but rescaling $B = \lambda(e_1 + e_2 + e_3)$ and $D = \lambda(e_1 + e_2 - e_3)$ then one can determine the scale factor $\lambda$ by the requirement that $AB$ and $BD$ are orthogonal to the vertex $q$. This will determine the scale factor $\lambda = \frac{1}{2}$. One can easily check that the vertices $ABCD$ form a rhombus orthogonal to the vertex $q = e_1 + e_2$. This shows that the *rhombic dodecahedron* has 14 vertices lying on two orbits given by the set of quaternions $O(100) = \{\pm e_1, \pm e_2, \pm e_3\}$ and the set $\frac{1}{\sqrt{2}} O(001) = \frac{1}{2}(\pm e_1 \pm e_2 \pm e_3)$. It is depicted in the Figure 2(b). The system can be rotated by the element $[\frac{1}{2}(1 + e_1 + e_2 + e_3), \frac{1}{2}(1 - e_1 - e_2 - e_3)]$ of the group $W(B_3)$ to determine the vertex $q' = e_2 + e_3$ where $A \to C = e_2$, $C \to E = e_3$, $B$ remains fixed and $D \to F = \frac{1}{2}(-e_1 + e_2 + e_3)$. The new rhombus $CBEF$ is orthogonal to the vertex $q'$. The dihedral angle then is $\theta = 120^0$. Two orbits lie on concentric spheres of radii $r_1 = 1$ and $r_2 = \frac{\sqrt{3}}{2} \approx 0.866$. The *rhombic dodecahedron* has 14 vertices 12 faces (rhombus) and 24 edges. The centers of edges are represented by the set of 24 quaternions

$$\frac{1}{2}(\pm e_1 \pm e_2 \pm 3e_3), \frac{1}{2}(\pm e_1 \pm 3e_2 \pm e_3), \frac{1}{2}(\pm 3e_1 \pm e_2 \pm e_3). \tag{19}$$

The set of quaternions in (19) represents the vertices of the union of two truncated tetrahedra, one defined by the orbit $O(110)$ as given in (13) and the other by the orbit $O(011)$ which is obtained by changing the sign of the elements of the orbit $O(110)$ of (13). Since the octahedral group is $W(B_3) \approx Aut(A_3)$, the set of quaternions in (19) is a single orbit under the octahedral group. Therefore the *rhombic dodecahedron* is not only face transitive but also the edge transitive. Since we have $\frac{|W(B_3)|}{|O(010)|} = 4$, the symmetry of the face is a group of order 4. The only symmetry in this case is the group $C_2 \times C_2$ generated by the elements $[e_3, -e_3]^*$ and $[\frac{1}{\sqrt{2}}(e_1 - e_2), -\frac{1}{\sqrt{2}}(e_1 - e_2)]^*$ which fixes the rhombus $ABCD$.

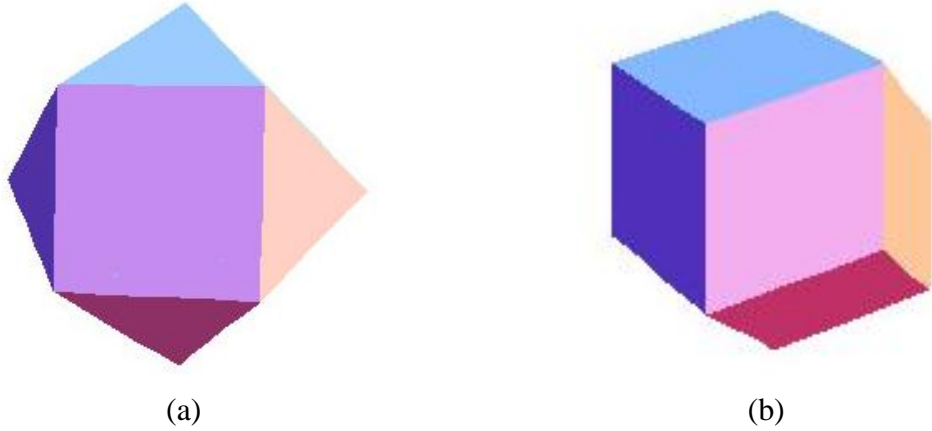

(a)          (b)

Figure 2. The cuboctahedron (a) and its dual rhombic dodecahedron (b)

*3.2*          *Tetrakis Hexahedron* ( dual of truncated octahedron)

Vertices of the truncated octahedron can be determined from the highest weight $\Lambda = (110) = (100) + (010) = e_1 + (e_1 + e_2) = 2e_1 + e_2$. Truncated octahedron is similar to the cuboctahedron however triangular faces are replaced by the hexagonal faces. As in the previous case, the centers of the 6 square faces are represented by the elements of the orbit $O(100) = \{\pm e_1, \pm e_2, \pm e_3\}$ and the centers of the 8 hexagonal faces are represented by the elements of the orbit $O(001) = \frac{1}{\sqrt{2}}(\pm e_1 \pm e_2 \pm e_3)$ up to some scale factors. Two hexagonal faces and one square face meet at a common vertex as shown in Figure 3(a). If we take the common vertex to be the quaternion $q = 2e_1 + e_2$ then the surrounding vertices of the isosceles triangle can be taken to be $A = \lambda e_1$, B=$e_1 + e_2 - e_3$, C=$e_1 + e_2 + e_3$. It is obvious that the line $BC$ is orthogonal to the vertex $q$. We check that $q$ will be orthogonal to the plane of the isosceles triangle $ABC$ if $\lambda = \frac{3}{2}$. The action of the group $W(B_3)$ on these quaternions will generate the 14 vertices of the of the t*etrakis hexahedron*. This proves that the set of 8 vertices lie on a sphere of radius $r_1 = \sqrt{3} \approx 1.73$ and the 6 vertices lie on the sphere of radius $r_2 = 1.5$. The symmetry of the isosceles triangle $ABC$ is the reflection group $C_2$ with respect to the $x-y$ plane generated by $[e_3, -e_3]^*$.

The number of edges can be easily determined as $6 \times 4 + \frac{8 \times 3}{2} = 36$. A rotation by the element $[\frac{1}{\sqrt{2}}(1+e_1), \frac{1}{\sqrt{2}}(1-e_1)]$ around the unit quaternion $e_1$ by $\pi/2$ will lead to a new vertex $q' = 2e_1 + e_3$ and the triangle $ACD$ with $D = e_1 - e_2 + e_3$. The dihedral angle between two adjacent triangles is $\theta = 143°7'48''$. The faces of the t*etrakis hexahedron* are represented by the 24 vertices of the truncated octahedron and are given as

$$(\pm 2e_1 \pm e_2), (\pm 2e_2 \pm e_3), (\pm 2e_3 \pm e_1), \quad (\pm e_1 \pm 2e_2), (\pm e_2 \pm 2e_3), (\pm e_3 \pm 2e_1). \qquad (20)$$

The sets of vertices of the t*etrakis hexahedron* consist of two orbits of $W(B_3)$ and are given by

$$\{\pm \frac{3}{2}e_1, \pm \frac{3}{2}e_2, \pm \frac{3}{2}e_3\}, \qquad (\pm e_1 \pm e_2 \pm e_3). \qquad (21)$$

The t*etrakis hexahedron* with the above vertices are illustrated in Figure 3(b)

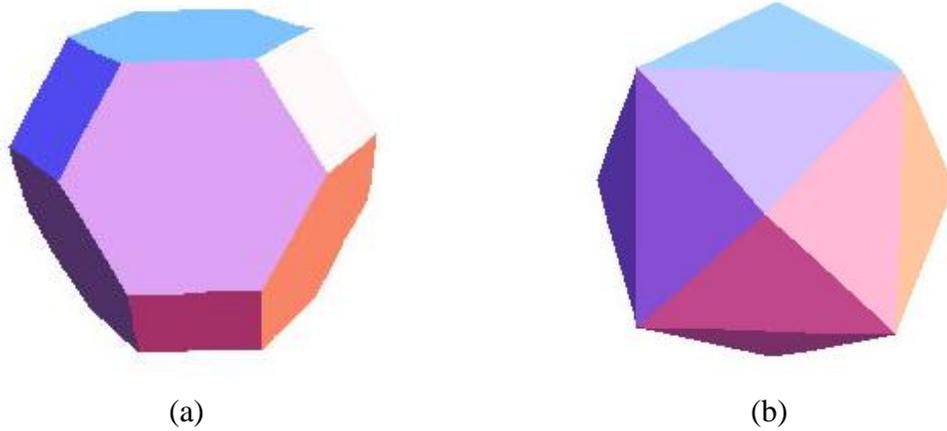

(a)          (b)

Figure 3. The truncated octahedron(a) and its dual t*etrakis hexahedron*(b)

*3.3 Triakis Octahedron* ( dual of truncated cube)

As shown in Figure 4(a) the truncated cube consists of 6 octagonal and 8 triangular faces. While the centers of the octagonal faces are represented by the set of vertices of octahedron up to a scale factor and the centers of the triangular faces are represented by the set of quaternions corresponding to the vertices of a cube subject to a change of scale. Let the vertex $\Lambda = (011) = q_3 = e_1 + e_2 + \frac{1}{\sqrt{2}}(e_1 + e_2 + e_3)$ represents the quaternion orthogonal to the isosceles triangle $ABC$ represented by the quaternions $A = \lambda(e_1 + e_2 + e_3)$, $B = e_1$ and $C = e_2$. The line $BC$ is orthogonal to $q_3$ and $\lambda$ is determined as $\lambda = \sqrt{2} - 1$ from the requirement that the triangle $ABC$ be orthogonal to $q_3$. Then the set of vertices of *triakis octahedron* are given by 14 quaternions

$$(\sqrt{2}-1)(\pm e_1 \pm e_2 \pm e_3), \qquad \{\pm e_1, \pm e_2, \pm e_3\}. \tag{22}$$

This means that the 8 vertices of *triakis octahedron* are on a sphere $r_1 = (\sqrt{2}-1)\sqrt{3} \approx 0.717$ and the 6 vertices are on a sphere of radius 1. It is shown in Figure 4(b). The symmetry of the above triangle $ABC$ is the cyclic group $C_2$ generated by the element $[\frac{1}{\sqrt{2}}(e_1 - e_2), -\frac{1}{\sqrt{2}}(e_1 - e_2)]^*$ which interchanges $e_1 \leftrightarrow e_2$ and fixes $e_3$. The vertices of the truncated cube which are orthogonal to the faces of the *triakis octahedron* are given by the set of 24 quaternions

$$(1+\frac{1}{\sqrt{2}})(\pm e_1 \pm e_2) \pm \frac{1}{\sqrt{2}} e_3; (1+\frac{1}{\sqrt{2}})(\pm e_2 \pm e_3) \pm \frac{1}{\sqrt{2}} e_1;$$
$$(1+\frac{1}{\sqrt{2}})(\pm e_3 \pm e_1) \pm \frac{1}{\sqrt{2}} e_2 \tag{23}$$

The number of edges can be easily determined as $8 \times 3 + \frac{6 \times 4}{2} = 36$.

Rotating the system around the unit quaternion $\frac{1}{\sqrt{3}}(e_1 + e_2 + e_3)$ the triangle $ABC$ is rotated to the triangle $ACD$ where $D = e_3$ and the vertex $q_3$ is transformed to $q_1 = \frac{1}{\sqrt{2}} e_1 + (1+\frac{1}{\sqrt{2}})(e_2 + e_3)$. The dihedral angle between the two triangular faces is $\theta = 147^0 21'(0.4)''$.

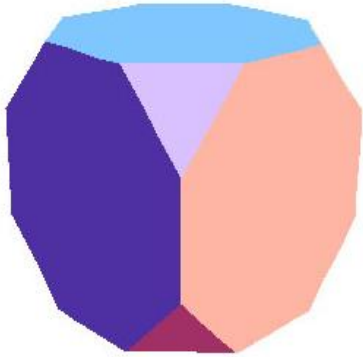
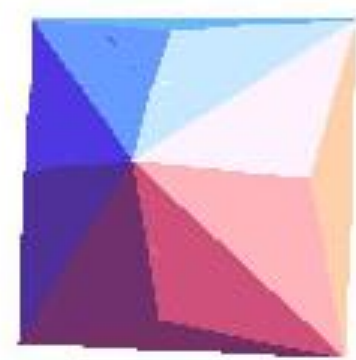

(a)          (b)

Figure 4. The truncated cube(a) and its dual triakis octahedron(b)

*3.4*           *Deltoidal icositetrahedron* ( dual of the small rhombicuboctahedron)

The small rhombicuboctahedron is displaced in Figure 5(a) which consists of 24 vertices, 48 edges and 26 faces (8 equilateral faces, 12+6 squares). It is clear from the classification of faces that the dual solid *deltoidal icositetrahedron* will consists of vertices which will be represented, up to scale factors, by three classes of orbits, namely, $O(001) = \frac{1}{\sqrt{2}}(\pm e_1 \pm e_2 \pm e_3)$, $O(010) = \{(\pm e_1 \pm e_2), (\pm e_2 \pm e_3), (\pm e_3 \pm e_1)\}$ and $O(100) = (\pm e_1, \pm e_2, \pm e_3)$. The vertex corresponding to the highest weight of the orbit of the small rhombicuboctahedron can be given by $q_1 = \Lambda = (101) = e_1 + \frac{1}{\sqrt{2}}(e_1 + e_2 + e_3)$. We have 24 vertices of the rhombicuboctahedron derived from $O(101) = W(B_3)(101)$:

$$\pm \alpha e_1 + \beta(\pm e_2 \pm e_3),\ \pm \alpha e_2 + \beta(\pm e_3 \pm e_1),\ \pm \alpha e_3 + \beta(\pm e_1 \pm e_2),$$
$$\alpha = (1 + \frac{1}{\sqrt{2}}),\ \beta = \frac{1}{\sqrt{2}}\ . \tag{24}$$

In small rhombicuboctahedron a vertex is surrounded by three squares and one equilateral triangle. It is easy to see that the centers of the squares and the triangle surrounding the vertex $q_1$ given above can be, up to the scale factors, given by the quaternions $A = e_1$, $B = \frac{1}{\sqrt{2}}(e_1 + e_2)$, $C = \lambda(e_1 + e_2 + e_3)$ and $D = \frac{1}{\sqrt{2}}(e_1 + e_3)$. Note that the lines *AB*, AD, BD are all orthogonal to the quaternion $q_1$. To have the vertex *C* in the same plane we should have $\lambda = \frac{\sqrt{2}+1}{\sqrt{2}+3} \approx 0.547$. We obtain a kite with $AB = AD$ and $CB = CD$. Rotating the system around the vertex *C* by $120^0$ one obtains the vertices $A' = e_2, B' = \frac{1}{\sqrt{2}}(e_2 + e_3), D' = B$ and $C' = C$ so that the *BC* is the common edge between two kites. The new kite $A'B'CB$ is orthogonal to the quaternion $q_2 = e_2 + \frac{1}{\sqrt{2}}(e_1 + e_2 + e_3)$ therefore the dihedral angle between two faces is $\theta = 138^07'5''$. The kite *ABCD* has the symmetry group $C_2$ generated by the element $[\frac{1}{\sqrt{2}}(e_1 - e_2), -\frac{1}{\sqrt{2}}(e_1 - e_2)]^*$. The 26 vertices of the *deltoidal icositetrahedron* are given by the sets of quaternions

$$\{\pm e_1, \pm e_2, \pm e_3\},\ \{\frac{1}{\sqrt{2}}(\pm e_1 \pm e_2), \frac{1}{\sqrt{2}}(\pm e_2 \pm e_3), \frac{1}{\sqrt{2}}(\pm e_3 \pm e_1)\},\ \frac{\sqrt{2}+1}{\sqrt{2}+3}(\pm e_1 \pm e_2 \pm e_3). \tag{25}$$

It is shown in the Figure 5(b). It is clear from the discussion that the 18=6+12 vertices lie on a sphere of unit length and the remaining 8 vertices lie on a sphere of radius 0.547.

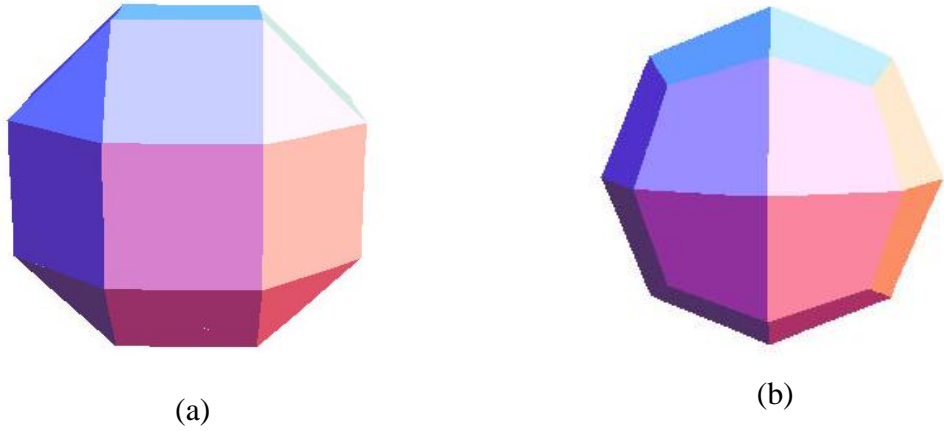

Figure 5. The small rhombicuboctahedron (a) and its dual deltoidal icositetrahedron

*3.5*  *Disdyakis dodecahedron* ( dual of the great rhombicuboctahedron)

The great rhombicuboctahedron has 6 octagonal faces, 8 hexagonal faces and 12 square faces as shown in the Figure 6(a). Since the great rhombicuboctahedron has 48 vertices the symmetry which fixes one vertex is trivial, namely just the unit element. This means the face of the d*isdyakis dodecahedron* has no symmetry at all. The face will be a scalene triangle as we discuss now. The highest weight for the great rhombicuboctahedron is

$$q_1 = \Lambda = (111) = e_1 + (e_1 + e_2) + \frac{1}{\sqrt{2}}(e_1 + e_2 + e_3) = \alpha e_1 + \beta e_2 + \gamma e_3 \qquad (26)$$

where $\alpha = 2 + \frac{1}{\sqrt{2}},\ \beta = 1 + \frac{1}{\sqrt{2}},\ \gamma = \frac{1}{\sqrt{2}}$. This vertex is surrounded by one octagon, one hexagon and one square, centers of which, are represented respectively by the quaternions $A = e_1$, $B = \lambda(e_1 + e_2)$ and $C = \eta(e_1 + e_2 + e_3)$. The parameters $\lambda$ and $\eta$ are determined from the requirement that the vertex $q_1$ be orthogonal to the triangle $ABC$ as $\lambda = \frac{2\sqrt{2}+1}{3\sqrt{2}+2} \approx 0.613$ and $\eta = \frac{2\sqrt{2}+1}{3\sqrt{2}+3} \approx 0.707$. Therefore the vertices of d*isdyakis dodecahedron* lie on three different concentric circles of radii $r_1 = 1$ (6 vertices), $r_2 = \sqrt{3}\eta \approx 0.916$ (8 vertices) and $r_3 = \sqrt{2}\lambda \approx 0.867$ (12 vertices). The 26 vertices can be determined by rescaling the orbits as $O(100), \lambda O(010)$ and $\sqrt{2}\eta O(001)$. It is plotted in Figure 6(b). To determine the adjacent triangle and the quaternion orthogonal to it can be achieved by applying the group element $[e_3, -e_3]^*$ on the vertices of the triangle $ABC$ to obtain the triangle $ABC'$ where $C' = \eta(e_1 + e_2 - e_3)$. The same group element transforms $q_1$ to $q_2 = e_1 + (e_1 + e_2) + \frac{1}{\sqrt{2}}(e_1 + e_2 - e_3)$. Then the dihedral angle between the faces is $\theta = 155^0 4' 56''$.

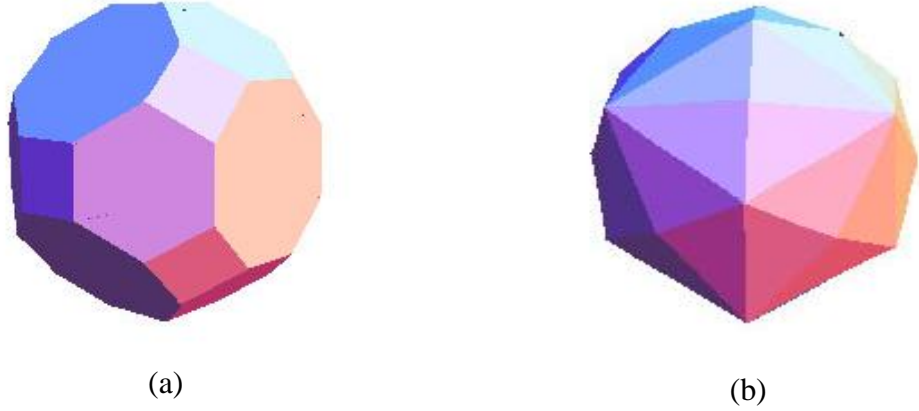

(a)                      (b)

Figure 6. The great rhombicuboctahedron (a) and its dual disdyakis dodecahedron

## 4.      The Catalan solids possessing the icosahedral symmetry group $W(H_3)$

We have discussed the Archimedean solids in reference [11] possessing the icosahedral symmetry $W(H_3)$ defined in (5). The orbits $O(\sigma 00)$ and $O(00\sigma)$ represent two dual Platonic solids, namely, dodecahedron and icosahedron respectively. Here $\sigma = \dfrac{1-\sqrt{5}}{2}$ and the golden ratio $\tau = \dfrac{1+\sqrt{5}}{2}$ satisfy the relations $\tau + \sigma = 1, \tau\sigma = -1, \tau^2 = \tau + 1, \sigma^2 = \sigma + 1$. An overall factor $\sigma$ comes from the projection of the 6-dimensional Dynkin–Coxeter root system of $D_6$ to 3-dimensions. Half of the 60 roots of $D_6$ represent the vertices o the icosidodecahedron and the remaining 30 is its scaled copy by $\sigma$. The Archimedean solids of the icosahedral symmetry are the orbits $O(0\sigma 0)$, $O(\sigma\sigma 0)$, $O(0\sigma\sigma)$, $O(\sigma 0\sigma)$ and $O(\sigma\sigma\sigma)$. They describe respectively the Archimedean solids, icosidodecahedron, truncated dodecahedron, truncated icosahedron, small rhombicosidodecahedron, and great rhombicosidodecahedron. Now we construct their Catalan solids in turn.

*4.1    Rhombic Triacontahedron* ( dual of the icosidodecahedron)

The orbit describing the vertices of the icosidodecahedron is given by [11]

$$O(0\sigma 0) = W(H_3)(0\sigma 0) =$$

$$\{ \pm e_1, \pm e_2, \pm e_3, \tfrac{1}{2}(\pm e_1 \pm \sigma e_2 \tau \pm e_3), \tfrac{1}{2}(\pm e_2 \pm \sigma e_3 \tau \pm e_1), \tfrac{1}{2}(\pm e_3 \pm \sigma e_1 \tau \pm e_2) \}. \tag{27}$$

Faces of the icosidodecahedron consist of 20 equilateral triangles and 12 regular pentagons as shown in the Figure 7(a). The *rhombic triacontahedron* has then 32=20+12 vertices and 30 faces. Order of the symmetry group of the faces, in other words, the order of the group fixing one vertex of the icosidodecahedron is $|W(H_3)|/30 = 4$. As it is expected, the face is a rhombus having the symmetry group $C_2 \times C_2$. Let $e_1$ represents the vertex, a common point of two pentagons and two triangles. One can easily determine the centers of the triangular faces as $A = \frac{\tau}{3}(\tau e_1 + \sigma e_2)$, $C = \frac{\tau}{3}(\tau e_1 - \sigma e_2)$ and the centers of the pentagonal faces as the quaternions $B = \frac{\tau+2}{5}(e_1 + \sigma e_3)$, $D = \frac{\tau+2}{5}(e_1 - \sigma e_3)$. These vertices define a golden rhombus meaning that the ratio of the diagonals is $\tau$. The rhombus $ABCD$ is orthogonal to the quaternion $e_1$. The centers of the triangular faces belong to the orbit

$$O(\sigma 00) = \{\frac{1}{2}(\pm \sigma e_1 \pm \tau e_3), \frac{1}{2}(\pm \sigma e_2 \pm \tau e_1), \frac{1}{2}(\pm \sigma e_3 \pm \tau e_2), \frac{1}{2}(\pm e_1 \pm e_2 \pm e_3)\}. \tag{28}$$

The centers of the pentagons belong to the orbit $O(00\sigma)$ given by

$$O(00\sigma) = \{\frac{1}{2}(\pm e_1 \pm \sigma e_3), \frac{1}{2}(\pm e_2 \pm \sigma e_1), \frac{1}{2}(\pm e_3 \pm \sigma e_2)\}. \tag{29}$$

Therefore the vertices of the *rhombic triacontahedron* consist of 32 vertices of (28-29) up to the scale factors defined above. This indicates that the 20 vertices lie on a sphere with the radius $|A| = \frac{\tau}{\sqrt{3}}$ and the 12 vertices lie on a sphere of radius $|B| = \frac{1}{\sqrt{\sigma+2}}$. Ratio between the two radii is $\frac{|A|}{|B|} = \sqrt{\frac{\tau+2}{3}} \approx 1.098$. If we rotate the system around the vertex $A$ which can be done by the group element $[q, \bar{q}]$ with $q = \frac{1}{2}(1 + \tau e_1 + \sigma e_2)$, then $e_1 \to \frac{1}{2}(\tau e_1 - e_2 - \sigma e_3)$ and the rhombus $ABCD$ is transformed to the adjacent one which is orthogonal to the quaternion $\frac{1}{2}(\tau e_1 - e_2 - \sigma e_3)$. Then one can determine the dihedral angle between two adjacent faces as $\theta = 144^0$. The Catalan solid, *rhombic triacontahedron* is not only face transitive but also edge transitive. The centers of the edges of the rhombus $ABCD$ are given by the set of quaternions

$$\frac{A+B}{2} = \frac{\tau}{6}(2\tau e_1 + \sigma e_2 - e_3), \quad \frac{A+D}{2} = \frac{\tau}{6}(2\tau e_1 + \sigma e_2 + e_3),$$
$$\frac{C+B}{2} = \frac{\tau}{6}(2\tau e_1 - \sigma e_2 - e_3), \quad \frac{C+D}{2} = \frac{\tau}{6}(2\tau e_1 - \sigma e_2 + e_3) \tag{30}$$

These vertices belong to the orbit of size 60 edges obtained from the highest weight $\sigma(10\tau) = \frac{1}{2}(-\sigma e_1 + e_2 + 2\tau e_3)$. The 60 quaternions representing the centers of the edges can be obtained by applying $W(H_3)$ on any quaternion in (30) or on the highest weight. In short, the *rhombic triacontahedron* has 30 faces, 32 vertices and 60 edges.

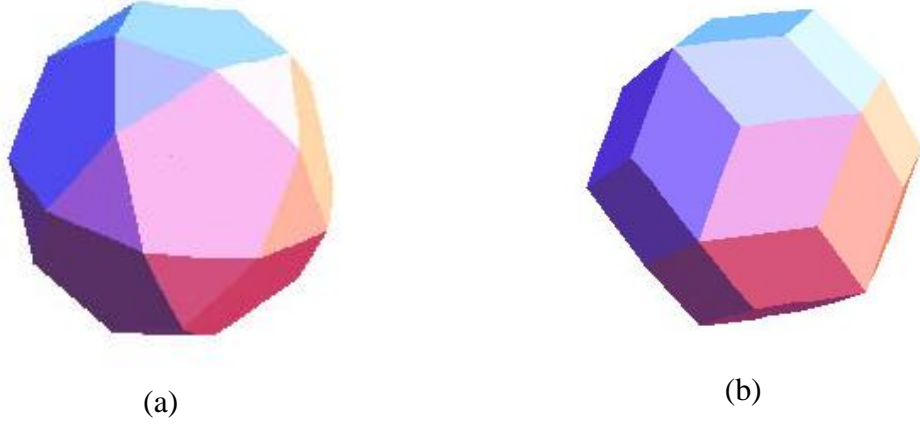

Figure 7. Icosidodecahedron (a) and its dual rhombic triacontahedron (b)

*4.2 Triakis Icosahedron* (dual of the truncated dodecahedron)

The truncated dodecahedron has 20 triangular and 12 decagonal faces with 60 vertices as well as 90 edges as shown in Figure 8(a). Therefore the dual solid will have 32 vertices consisting of two orbits and 60 faces. From the symmetry consideration, the order of the group fixing one vertex of the truncated dodecahedron is $|W(H_3)|/60 = 2$. This indicates that the face of the dual solid has a face of cyclic symmetry $C_2$ implying that the face is an isosceles triangle as we discuss now.

The highest weight of the truncated dodecahedral orbit is $q_1 = \Lambda = \sigma(110) = \frac{1}{2}(-\sigma e_1 + \tau e_3) + e_3$ which is the sum of the highest weights describing dodecahedron and the icosidodecahedron. The centers of two decagonal faces to this vertex can be taken as $B = -\sigma e_2 + e_3$ and $C = \sigma e_2 + e_3$ where $|B| = |C|$ and the line $BC$ is orthogonal to the vertex $q_1$ represented by the highest weight. The vertex $A = \lambda(-\sigma e_1 + \tau e_3)$ which represents the center of the triangular face, up to a scale factor, determines the isosceles triangle $ABC$ orthogonal to the vertex $q_1$ provided $\lambda = \frac{\tau + 2}{2\tau + 3}$ which leads to the ratio of two radii of two concentric spheres $\frac{|A|}{|B|} = \frac{\tau + 2}{2\tau + 3}\sqrt{\frac{3}{\sigma + 2}} \approx 0.855$. This shows that in the outer sphere we have 12 icosahedral vertices, one of which is connected to two vertices in the inner sphere which represent the dodecahedral vertices. They are given in (28-29). When the system is rotated by $120^0$ around the vertex $A$ one obtains an equilateral triangle $BCD$ where $A$ is the center of the triangle $BCD$ up to a scale factor which divides the equilateral triangle into three isosceles triangles whose normal vectors are represented by the quaternions

$$q_1 = \frac{1}{2}(-\sigma e_1 + (\tau + 2)e_3), \quad q_2 = \frac{1}{2}(\tau e_1 + \sigma e_2 + 2\tau e_3), \quad q_3 = \frac{1}{2}(\tau e_1 - \sigma e_2 + 2\tau e_3). \qquad (31)$$

The dihedral angle can be determined as $\theta = 160^0 36' 45''$ using the scalar product between any pair of vectors. The *triakis icosahedron* has been shown in Figure 8(b) where three isosceles triangles meet at one vertex.

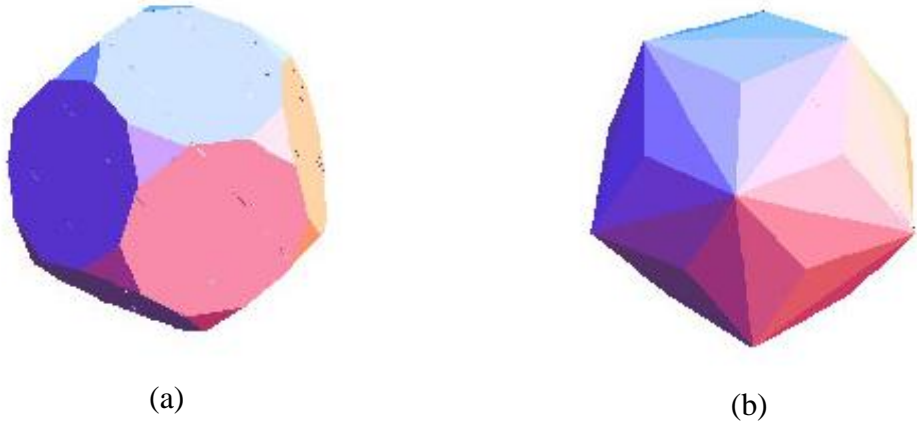

(a)            (b)

Figure 8. Truncated dodecahedron(a) and its dual solid triakis icosahedron(b)

*4.3 Pentakis Dodecahedron* (dual of the truncated icosahedron)

Truncated isosahedron with its 60 vertices, 32 faces (12 regular pentagons and 20 regular hexagons) is a model of $C_{60}$ molecule. Its dual, the Catalan solid p*entakis dodecahedron*, has 32 vertices (12 from the icosahedral orbit and 20 from the dodecahedral orbit) and 60 faces of isosceles triangles as expected from the fact that $|W(H_3)|/60 = 2$. The isosceles triangle, determined by the vertices $A = \lambda(-\sigma e_2 + e_3)$, B=$(-\sigma e_1 + \tau e_3)$, C=$(e_1 + e_2 + e_3)$ is orthogonal to the vertex $q = \frac{1}{2}(e_1 - 2\sigma e_2 + \tau^2 e_3)$ which determines the factor $\lambda = \frac{3\tau}{\sigma + 4}$. The ratio of radii of two spheres having 20 and 12 vertices respectively is $\frac{|B|}{|A|} = \frac{\sigma + 4}{\sqrt{3(\tau + 2)}} \approx 1.0265$. This shows that the dodecahedral vertices lie on the outer sphere and the icosahedral vertices lie on the inner sphere. Five isosceles triangles meet at one vertex of the icosahedral vertices and six of them meet at one dodecahedral vertices. The dihedral angle between two adjacent faces can be determined as $\theta = 156^0 43' 7''$ in a similar manner as explained in other cases. To count the number of edges is also easy. From each icosahedral vertex 5 equal edges originate yielding $12 \times 5 = 60$ and for each triangle there exits one more extra edge shared by another triangle which is $\frac{60}{2} = 30$. Therefore a total number of edge is 90. The whole set of 32 vertices of the p*entakis dodecahedron* can be written as

$$\{ (\pm\sigma e_1 \pm \tau e_3), (\pm\sigma e_2 \pm \tau e_1), (\pm\sigma e_3 \pm \tau e_2), (\pm e_1 \pm e_2 \pm e_3) \},$$

$$\frac{3\tau}{\sigma + 4}\{(\pm e_1 \pm \sigma e_3), (\pm e_2 \pm \sigma e_1), (\pm e_3 \pm \sigma e_2)\} . \tag{32}$$

The truncated icosahedron as well as its dual p*entakis dodecahedron* are displayed in Figure 9.

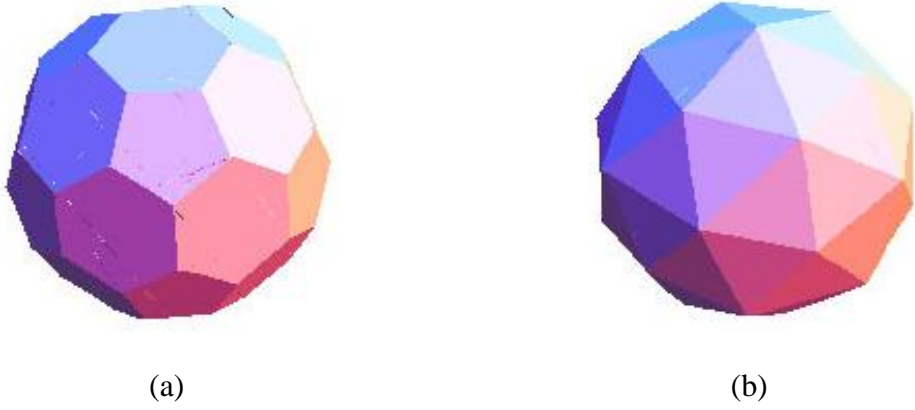

(a)          (b)

Figure 9. Truncated icosahedron(a) and its dual p*entakis dodecahedron*(b)

*4.4 Deltoidal Hexacontahedron* (dual of small rhombicosidodecahedron)

The small rhombicosidodecahedron consists of 60 vertices, 62 faces (12 pentagons, 30 squares and 20 equilateral triangles) and 120 edges as shown in the Figure 10(a).The Catalan solid, *deltoidal hexacontahedron*, as expected, consists of 62 vertices which will be in 3 sets of icosahedral, dodecahedral and icosidodecahedral orbits. One vertex of the small rhombicosidodecahedron is shared by one pentagonal, one triangular and two square faces. The symmetry fixing one vertex of the small rhombicosidodecahedron which is a cyclic group $C_2$. It indicates that the face of the *deltoidal hexacontahedron* will be a kite of two-fold symmetry. The highest weight leading to the orbit of vertices of the small rhombicosidodecahedron is $q = \Lambda = \sigma(101) = \frac{1}{2}(-\sigma e_1 - \sigma e_2 + \tau^2 e_3)$. The vertices of the kite which is orthogonal to this vertex is determined to be

$$A = \lambda(-\sigma e_2 + e_3), \ B = e_3, \ C = \eta(-\sigma e_1 + \tau e_3), \ D = \frac{1}{2}(e_1 - \sigma e_2 + \tau e_3) \ . \tag{33}$$

The orthogonality of the vertex $q$ to the kite $ABCD$ determines $\lambda = \frac{\tau^2}{3}$ and $\eta = \frac{\tau^2}{\tau + 3}$. Vertices of the *deltoidal hexacontahedron* lie on three spheres with the radii $|A| = \frac{\tau\sqrt{\tau + 2}}{3} \approx 1.0259$, $|B| = |D| = 1$, $|C| = \frac{\tau^2\sqrt{3}}{\tau + 3} \approx 0.982$. On the outer sphere there exit 12 icosahedral vertices, in the middle sphere, 30 icosidodecahedral vertices and in the inner sphere 20 dodecahedral vertices. A rotation around the vertex $A$ by $\frac{2\pi}{5}$ would take $q$ to $q' = \frac{1}{2}(e_1 + \tau e_2 + 2e_3)$. Then the scalar product of these two vectors determine the dihedral angle as $\theta = 154^0 7' 17''$. The 60 faces of the *deltoidal hexacontahedron* is determined by the orbit $O(\sigma 0 \sigma)$ which can be found in the reference [11]. It is shown in the Figure 10(b).

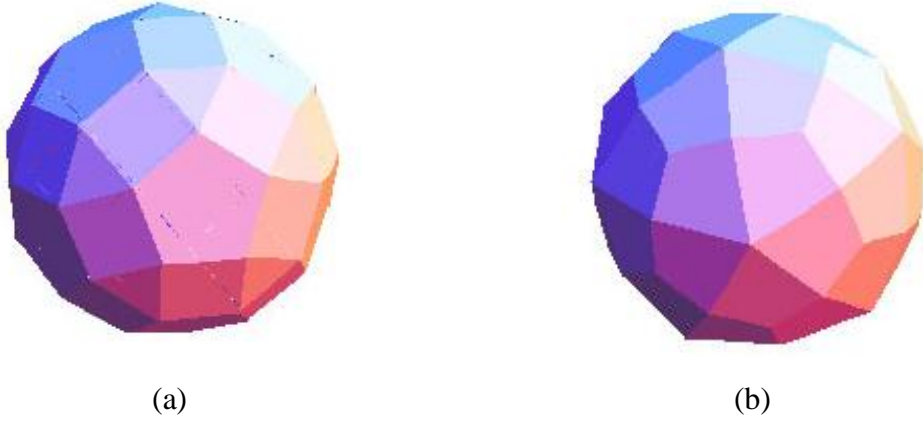

(a) (b)

Figure 10. The small rhombicosidodecahedron(a) and its dual *deltoidal hexacontahedron*(b)

*4.5 Disdyakis Triacontahedron* (dual of the great rhombicosidodecahedron)

Vertices of the great rhombicosidodecahedron can be obtained by applying $W(H_3)$ on what is called the highest weight

$$\Lambda = \sigma(111) = \sigma(100) + \sigma(010) + \sigma(001) = \frac{1}{2}(-\sigma e_1 + \tau e_3) + e_3 + \frac{1}{2}(-\sigma e_2 + e_3). \qquad (34)$$

The whole set of 120 vertices can be found in the reference [11]. The great rhombicosidodecahedron, as shown in the Figure 11(a), has 12 decagonal, 20 hexagonal and 30 square faces. At one vertex three different faces meet. It is then obvious that the face of the dual polytope *disdyakis triacontahedron* is a scalene triangle without any non-trivial symmetry. One can argue that the three vertices $A = \lambda(-\sigma e_2 + e_3)$, $B = \eta(-\sigma e_1 + \tau e_3)$, $C = \frac{1}{2}(e_1 - \sigma e_2 + \tau e_3)$ representing respectively the centers of the decagonal, hexagonal and the square faces form a triangle and it is orthogonal to the vertex $q = \frac{1}{2}(\tau e_1 - 2e_2 + \tau^3 e_3)$, which is one of the vertices of the great rhombicosidodecahedron. The factors are determined, as usual, from the orthogonality of the vertex $q$ to the triangle *ABC* which results in $\lambda = \frac{\tau + 3}{5}$ and $\eta = \frac{2\sigma + 3}{3}$. This shows that the vertices of the *disdyakis triacontahedron* lie on three concentric spheres. Let the outermost sphere has a radius $|A| = \frac{\tau + 3}{5}\sqrt{\sigma + 2} \approx 1.0858$ having 12 vertices of icosahedron, the next sphere contains 20 vertices of a dodecahedron with the radius $|B| = \frac{2\sigma+3}{\sqrt{3}} \approx 1.0184$ and the inner sphere $|C| = 1$ with 30 vertices of icosidodecahedron. When we rotate the system around the vertex $A$ by $\frac{2\pi}{5}$ one obtains the vertex $q' = \frac{1}{2}(2e_1 + \sqrt{5}e_2 + (\tau+2)e_3)$ which is orthogonal to the triangle $AB'C'$ where $B'$ and $C'$ are obtained from $B$ and $C$ respectively with the same rotation. The dihedral angle between two adjacent faces is then $\theta = 164^0 53' 16''$. The *disdyakis triacontahedron* is shown in the Figure 11(b).

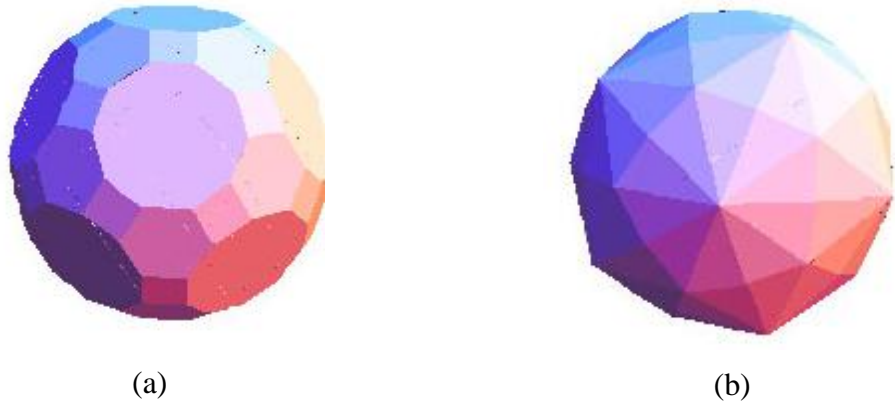

(a)          (b)

Figure11. The great rhombicosidodecahedron(a) and its dual *disdyakis triacontahedron*(b)

## 5. Conclusion

In this work we displayed a systematic construction of the Catalan solids, the dual solids of the Archimedean solids, with the use of Coxeter-Weyl groups $W(A_3), W(B_3), W(H_3)$. We employed the highest weight method for the irreducible representations of Lie algebras to determine the orbits. Catalan solids are face transitive meaning that the faces are transformed to each other by the Coxeter-Weyl group. The vectors orthogonal to the faces are the vertices of the Archimedean solids. The vertices of the Catalan solids are the unions of the orbits obtained from the fundamental weights. The vertices are on concentric spheres determined by the lengths of the fundamental weights up to some scale factors.

The Platonic solids and the Archimedean solids have been successfully applied to describe the crystallography in physics, molecular symmetries in chemistry and some virus structures in biology. In particular, the Coxeter group $W(H_3)$ representing the icosahedral symmetry with inversion in 3-dimensional Euclidean space is very useful in understanding the quasicrystallography in physics [16]. The polyhedra possessing the icosahedral symmetry have been successfully used in chemistry [17] and biology [18] for the symmetries of molecules and viral capsids which also requires the Catalan solids [19]. Therefore construction of the vertices of the polyhedra, through the Coxeter-Dynkin diagrams $A_3$, $B_3$, $H_3$ with quaternions, whether they are Platonic, Archimedean or Catalan solids will be very useful in the applications of the physical sciences displaying symmetries.